\documentclass[twocolumn,floatfix,superscriptaddress,longbibliography,nofootinbib]{revtex4-1}
\RequirePackage[sort&compress]{natbib}

\usepackage{multirow}
\usepackage{color}
\usepackage{soul}
\usepackage{tabularx}
\usepackage{amssymb,amsfonts,amsmath}
\usepackage{bm}
\usepackage[pdftex]{graphicx}
\usepackage{xcolor}
\usepackage{comment}
\usepackage{float}

\usepackage{supertabular}
\usepackage{longtable}

\usepackage[colorlinks = true,
            linkcolor = blue,
            urlcolor  = blue,
            citecolor = blue,
            anchorcolor = blue]{hyperref}

\usepackage{dcolumn}
\usepackage{bm}

\usepackage[caption=false]{subfig}
\usepackage{xcolor}

\renewcommand{\(}{\left(}
\renewcommand{\)}{\right)}
\renewcommand{\[}{\left[}

\newcommand{\tr}[1]{\text{Tr}\(#1\)}
\newcommand{\ket}[1]{|#1\rangle}
\newcommand{\bra}[1]{\langle#1|}
\newcommand{\braket}[2]{\langle#1|#2\rangle}

\usepackage{xcolor}
\definecolor{RoyalBlue}{HTML}{4169e1}
\definecolor{ForestGreen}{HTML}{228b22}

\usepackage{setspace}
\usepackage{comment}

\usepackage{todonotes}

\usepackage{tcolorbox}
\definecolor{airforceblue}{rgb}{0.36, 0.54, 0.66}
\definecolor{alizarin}{rgb}{0.82, 0.1, 0.26}

\begin{document}

\title{Generalized network density matrices for analysis of multiscale functional diversity}

\author{Arsham Ghavasieh}
\email[Corresponding author:~]{aghavasieh@fbk.eu}%
\affiliation{Fondazione Bruno Kessler, Via Sommarive 18, 38123 Povo, Italy}
\affiliation{Department of Physics, University of Trento, Via Sommarive 14, 38123 Povo (TN), Italy}
\author{Manlio De Domenico}
\email[Corresponding author:~]{manlio.dedomenico@unipd.it
}%
\affiliation{Department of Physics and Astronomy ``Galileo Galilei'', University of Padova, Padova, Italy}

\setlength {\marginparwidth }{2cm}

\date{\today}

\begin{abstract}
The network density matrix formalism allows for describing the dynamics of information on top of complex structures and it has been successfully used to analyze from system's robustness to perturbations to coarse graining multilayer networks from characterizing emergent network states to performing multiscale analysis. However, this framework is usually limited to diffusion dynamics on undirected networks. Here, to overcome some limitations, 
we propose an approach to derive density matrices based on dynamical systems and information theory, that allows for encapsulating a much wider range of linear and non-linear dynamics and richer classes of structure, such as directed and signed ones. We use our framework to study the response to local stochastic perturbations of synthetic and empirical networks, including neural systems consisting of excitatory and inhibitory links and gene-regulatory interactions. Our findings demonstrate that topological complexity does not lead, necessarily, to functional diversity---i.e., complex and heterogeneous response to stimuli or perturbations. Instead, functional diversity is a genuine emergent property which cannot be deduced from the knowledge of topological features such as heterogeneity, modularity, presence of asymmetries or dynamical properties of a system.

\end{abstract}

\maketitle

\section{Introduction}

Originally, density matrices have been introduced to represent quantum systems, in terms of probabilities of physical states and their quantum correlations~\cite{Fano_1957}. Half a century later, different attempts have been made to extend density matrices to classical complex systems~\cite{Severini_density_1,Severini_density_2,de2015reducibility,de2016spectral,ghavasieh2020statistical}, to capture the properties of interconnected nodes and their correlations, in a unifying framework. 

The density matrix capturing the statistical physics of complex information dynamics has found applications from centrality and robustness analysis~\cite{ghavasieh_Structural_robustness,ghavasieh_functional_robustness}, to identification of functional modules~\cite{ghavasieh_fungal} and classification of networks~\cite{de2016spectral,ghavasieh_SARSCOV2,ghavasieh_density_brain,topology_identification}(for a recent review, see Ref.~\cite{Ghavasieh_perspective}), from network phase transitions~\cite{Gabrielli_phase_transition} to renomarlization group~\cite{Gabrielli_RG_density}.

One reason for the broad applicability of this framework is that it is not limited to structural analysis--- i.e., not determined by the mere adjacency matrix: instead, it gives insights into the non-trivial coupling between the structure and dynamical processes~\cite{Vespignani_dynamics,Masuda_random_walk,dedomenico_spreading,lambiotte_markov,Estrada2008,Barzel_universality,Barzel_information_flow,Barzel_signal_propagation,Jesus_Gomez_exploding}--- and allows one to study the statistics of perturbation propagation at different scales (short- to long-range). More technically, to derive the density matrix, one solves a linear equation governed by a control operator $\hat{\mathbf{H}}$ that describes the dynamics of a field on top of the network, leading to a time-evolution matrix $\hat{\mathcal{G}}_\tau=e^{-\tau \hat{\mathbf{H}}}$ whose elements encode the flow of the field between the nodes and $\tau$ a fixed parameter encoding the propagation scale. The density matrix is obtained by normalizing the propagator by its trace, providing an ensemble that describes the statistics of the information dynamics on top of system at the given propagation scale. However, for non-hermitian control operators $\hat{\mathbf{H}}\neq\hat{\mathbf{H}}^{\dag}$, where $\hat{\mathbf{H}}^{\dag}$ is the complex conjugate of $\hat{\mathbf{H}}$, the probabilistic interpretation of the ensemble is difficult to reach, due to the presence of complex numbers in the spectrum. This limitation rules out a range of interesting systems with nonsymmetric structure and dynamics like reaction diffusion and synchronization. Furthermore, the framework assumes that the propagation starts from one of the nodes and assigns equal probabilities of being perturbed to each of them. While, many complex systems exhibit heterogeneity beyond such assumptions---e.g., in the connectome, the propagation of signals is more likely to start from sensory areas and less likely from the ones that process the sensory information~\cite{Sporns_sensorimotor_information_flow}. Here we provide a different formulation of density matrices, from the point of view of information theory, that completely resolves the aforementioned issues and greatly expands the range of applicability of the framework to nonlinear dynamics, including neural, gene-regulatory and epidemics, even on top of directed and signed networks.

\begin{figure*}
\centering
\includegraphics[width=.6\textwidth]{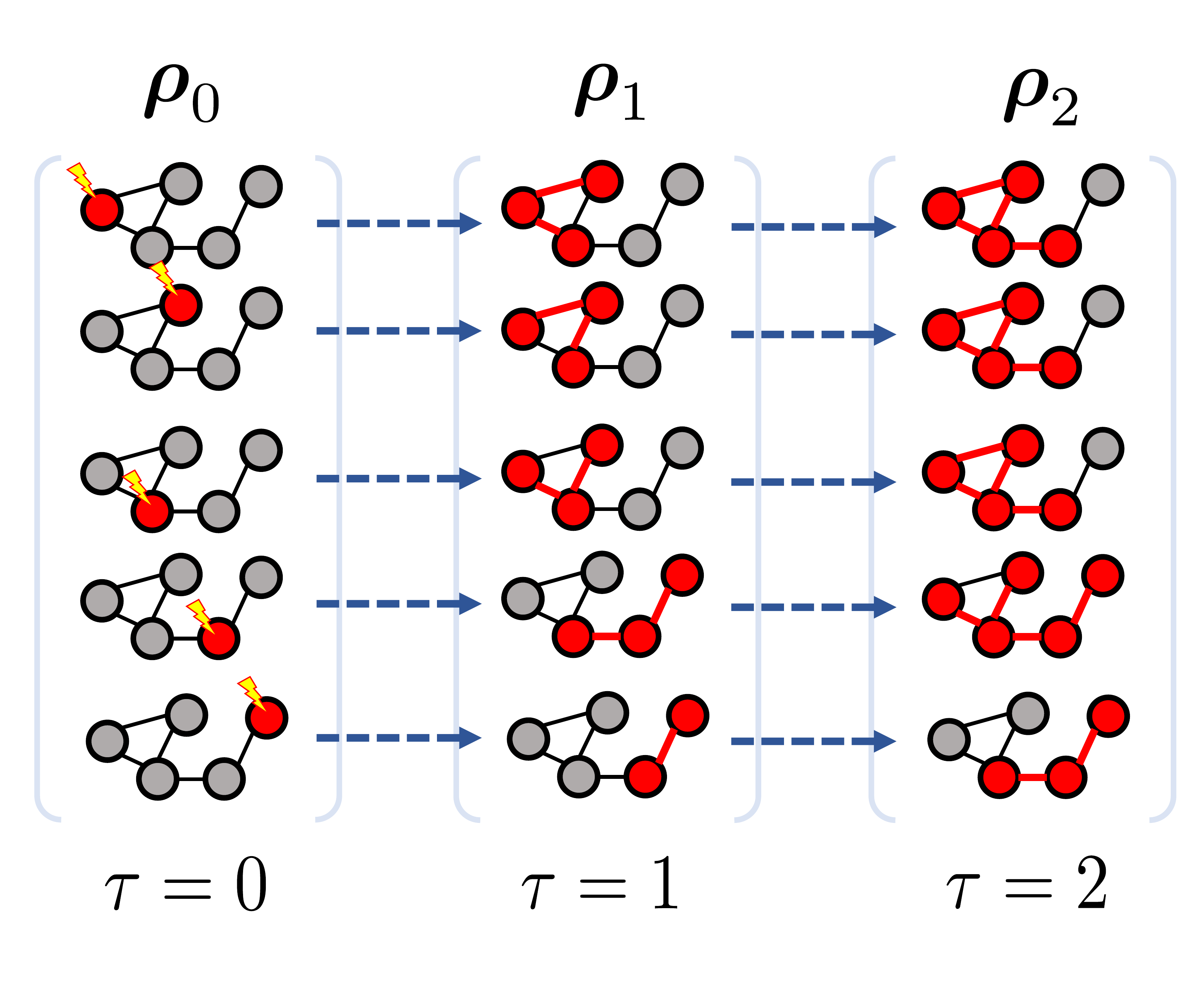}
\caption{\label{fig:schematic}\textbf{Density matrix as superposition of perturbed states.} A simple network of five nodes is considered here. A hypothetical discrete type of dynamics has been assumed for perturbation propagation, where red color means that the node is affected by the perturbation and red link indicates that the link has carried the perturbation between two nodes.
Each network depicted shows the system's response to a perturbation starting from a certain location and covering a specific propagation scale $\tau$. Fixing $\tau$, the density matrix encodes the average system's response to stochastic perturbations--- i.e., perturbations in different localities with certain probabilities.}

\end{figure*}

As an application, we focus on the functional diversity of biological systems and its fragility under structural damage. Functional diversity determines the range of possible dynamic responses of the system to environment or its parts and, as previously shown~\cite{ghavasieh2020statistical}, can be quantified by using the Von Neumann entropy of the density matrix. For this reason, we couple a range of synthetic and empirical networks with dynamical processes including neural and gene-regulatory, we calculate how perturbations of steady state propagate into the system in each case and, calculate the diversity of such propagation patterns in terms of the Von Neumann entropy. Interestingly, our finding clearly shows that functional diversity is an emergent property, which cannot be deduced from the knowledge of structural properties or dynamical rules of a system. The fact that at some propagation scales and for some dynamical configurations a random network provides the highest functional diversity, challenges the widely accepted assumption that a more complex topology guarantees a wider repertoire of response to internal and external stimuli. 

\section{Network density matrices.}

The coupling between networks and dynamical processes gives rise to the complex information dynamics observed in a multitude of biological systems. To model it and quantify its complexity, a field can be assumed on top of the network whose dynamics is governed by a linear differential equation with a control operator $\hat{\mathbf{H}}$ whose eigenvalues are denoted as $\lambda_{\ell}(\ell=1,2,...N$). The solution of such an equation is governed by a time-evolution operator $\hat{\mathcal{G}}_{\tau} = e^{-\tau \hat{\mathbf{H}}}$, where $\tau$ is the temporal parameter encoding the signal propagation scale. Eigen-decomposition of the time-evolution operator $\hat{\mathcal{G}}_{\tau}$ gives a set of stream operators $\{\hat{\mathbf{\sigma}}_{\ell}(\tau)\}$--- i.e., identified by the outer product of left and right eigenvectors of $\hat{\mathcal{G}}_{\tau}$---, guiding the flow of information, weighted by their contribution to the flow $e^{-\tau \lambda_{\ell}}$ which is the $\ell$-th eigenvalue of $\hat{\mathcal{G}}_{\tau}$. Using this information, it is possible to find a statistical description of the system, through a procedure similar to quantum statistical mechanics (See Appendices). The summation of the contributions defines the partition function of the system $Z_\tau = \sum\limits_{\ell=1}^{N}e^{-\tau \lambda_{\ell}}$ and the density matrix follows $\hat{\boldsymbol{\rho}}_\tau=\hat{\mathcal{G}}_{\tau}/Z_\tau$. Despite the success in analyzing a range of complex systems, this approach is limited for two reasons. Firstly, the eigenvalues of a valid network density matrix $\rho_{\ell}(\tau)=e^{-\tau \lambda_{\ell}}/Z_\tau$ are expected to be positive to encode the probabilities of activation of streams, requiring the control operator $\hat{\mathbf{H}}$ to be hermitian, ruling out a broad range of dynamical processes and limiting the analysis to diffusion dynamics on top of undirected and unsigned networks. Secondly, and related to the first point, the spectrum of $\hat{\mathbf{H}}$ must guarantee that metrics such as Von Neumann entropy and partition function derived from the density matrix are real and positive. As we show here, using mathematical treatments under specific conditions, the latter condition is satisfied for a wider variety of control operators (See Appendices). Yet, the first limitation presents a serious challenge, as a statistical ensemble having imaginary or negative probabilities is difficult to interpret from a physical perspective, in this case.

\section{Generalized network density matrices.} In the same spirit of the original density matrix formulation, we view the propagation of perturbations through a complex network as a model of information flow. However, here we use signal processing to understand how system's units communicate with each other. Accordingly, we build a density matrix that not only allows for considering linear dynamics with Hermitian and non-Hermitian control operators, but extends the applicability to non-linear dynamics, far from the steady state. To this aim, let's indicate the initial state of the field by $|\psi\rangle$ with $\braket{i}{\psi}$ indicating its value on top of node $i$ being a complex number. The initial state can represent any arbitrary distribution--- e.g., steady state $\partial_\tau |\psi\rangle=0$, the zero state $|\psi\rangle=0$, etc. We assume a local perturbation of size $\Delta_{i}$ on top of an arbitrary node $i$, shifting the initial state by $\Delta_i |i\rangle$. Here, the initial state would be updated to $|\psi^{(i)}_{0}\rangle = |\psi\rangle + \Delta_i |i\rangle$, where the perturbation vector is given by$|\Delta \psi^{(i)}_{0}\rangle = |\psi^{(i)}_{0}\rangle - |\psi\rangle = \Delta_i |i\rangle$. Note that, the perturbation can occur on top of multiple nodes simultaneously. For instance, assume a set of node $\zeta = \{i,j,k\}$ on top of them we can have the perturbations with sizes $\{\Delta_i,\Delta_j,\Delta_k\}$ and the initial state would be updated to $|\psi^{(\zeta)}_{0}\rangle = |\psi\rangle + \Delta_i |i\rangle+\Delta_j |j\rangle+\Delta_k |k\rangle$. Yet, we focus on perturbations on top of single nodes for simplicity. Depending on the specific dynamical rules of the system, the initial vector will evolve to $|\psi^{(i)}_{\tau} \rangle$, with the parameter $\tau$ indicating the temporal propagation scale of the signal and the upper index $(i)$ denoting that the location of the perturbation is on top of node $i$. Note that this derivation is valid for any type of dynamical evolution. The perturbation propagation vector is $|\Delta \psi^{(i)}_{\tau}\rangle = |\psi^{(i)}_{\tau}\rangle - |\psi \rangle$, and the propagation of perturbations from node $i$ to node $j$ is given by $\langle j |\Delta \psi^{(i)}_{\tau}\rangle$, at the propagation scale $\tau$. The vector $|\Delta \psi^{(i)}_{\tau}\rangle$ can be seen as system's response to a perturbation at the site $i$.

As mentioned above, propagation of perturbations from one node to another proxies their information exchange, suggesting an interpretation based on classical signal processing~\cite{Wang2022}: the signal amplitude from node $i$ to the node $j$ is $\langle j |\Delta \psi^{(i)}_{\tau}\rangle$, and the signal energy on top of node $j$ reads $\langle j |\Delta \psi^{(i)}_{\tau}\rangle\langle \Delta \psi^{(i)}_{\tau}|j\rangle = ||\langle j |\Delta \psi^{(i)}_{\tau}\rangle||^{2}$. Propagation of perturbations from node $i$ (or alternatively a set of nodes) can also be encoded in local propagators given by the outer product of propagation vector and its complex conjugate
\begin{equation}\label{eq:local_propagator}
    \hat{\mathbf{U}}^{(i)}_{\tau} = |\Delta \psi^{(i)}_{\tau}\rangle \langle \Delta \psi^{(i)}_{\tau}|,
\end{equation}
where the $j$-th diagonal element $\langle j|\hat{\mathbf{U}}^{(i)}_{\tau}|j\rangle$ gives the signal energy on top of node $j$, received from node $i$ at $\tau$. Also, the $jk$ off-diagonal element $\langle j|\hat{\mathbf{U}}^{(i)}_{\tau}|k\rangle$ encodes the covariance between node $j$ and $k$ in receiving signal amplitudes from $i$. Note that this notion of energy, borrowed from communication science and engineering, is compatible with physical energy in specific systems--- e.g., the energy of electromagnetic waves traveling between the nodes of a specific telecommunication network is related to the second power of absolute value of wave amplitude---, and totally different in others---e.g., traveling electrochemical signals in the human brain.

\begin{figure*}
\centering
\includegraphics[width=\textwidth]{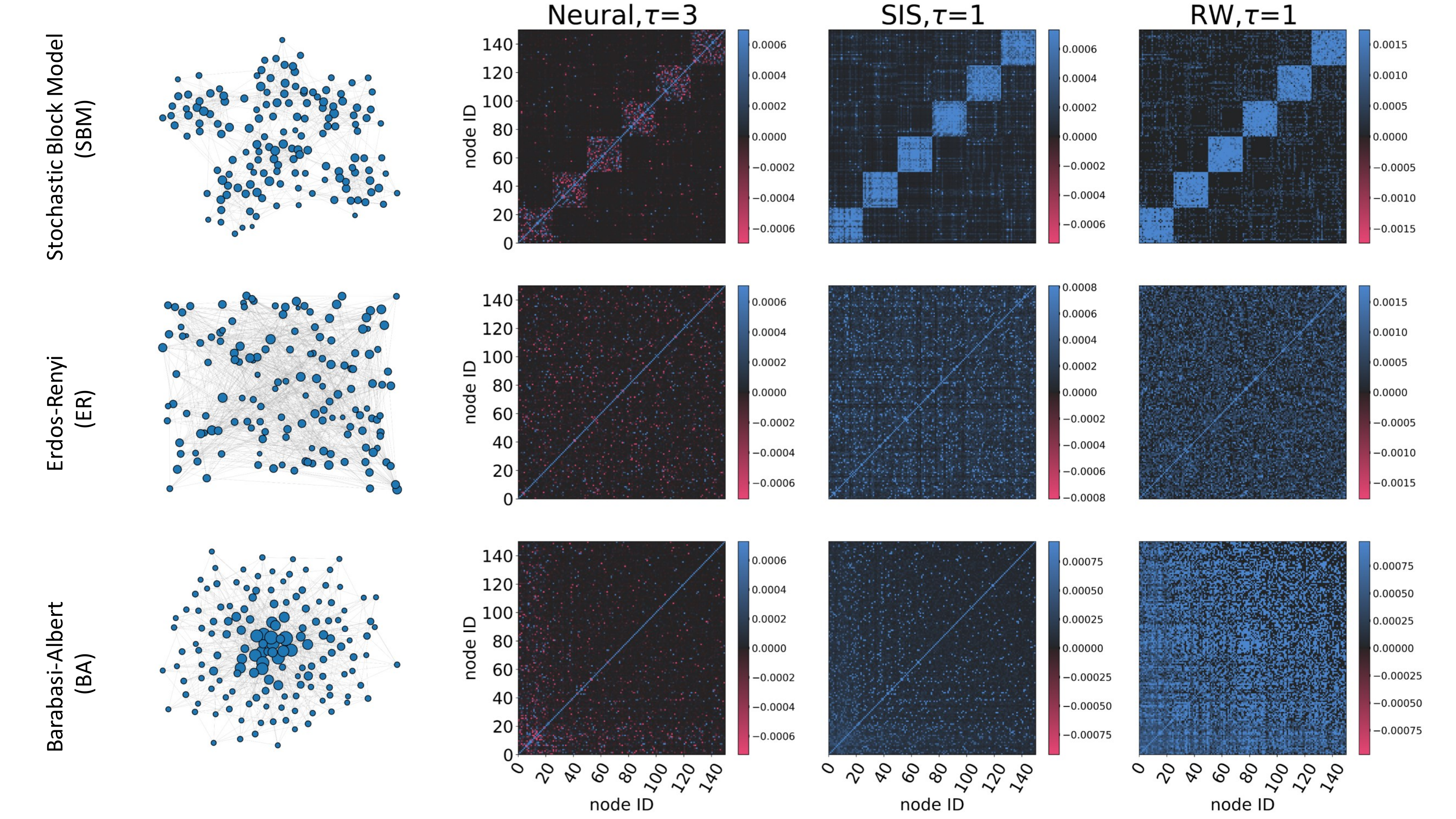}
\caption{\label{fig:static_networks_density}\textbf{Signal propagation on static networks: coupling structure and dynamics.} We show the density matrices for different synthetic systems, with structures given by Erdos-Renyi (ER), Barabasi-Albert (BA) and stochastic block model (SBM) and different dynamics: discrete random walk (RW), SIS epidemics and neural. The diagonal elements encode are signal energy on top of the nodes and the off-diagonal elements are the node-node covariance that can take positive or negative values depending on the dynamics and the structure being signed or not. Note that the diagonal elements are typically much larger than off-diagonal ones. Therefore, for clarity, the maximum value of the color spectrum of each heatmap corresponds to 10\% of the maximum value of the density matrix.}
\end{figure*}

Often, it is hard to precisely predict the location of perturbations. Therefore, we assume a distribution $p_{i},(i=1,2,...N)$ describing the probability of having a perturbation at each node. Consequently, system's perturbation propagation matrix can be encoded by a statistical propagator:
\begin{equation}\label{eq:statistical_propagator}
    \hat{\mathbf{U}}_{\tau} =\sum\limits_{i}p_{i}\hat{\mathbf{U}}^{(i)}_{\tau},
\end{equation}
where the $j$-th diagonal element $\langle j| \hat{\mathbf{U}}_\tau |j\rangle=\sum\limits_{i}p_{i}\langle j| \hat{\mathbf{U}}^{(i)}_\tau |j\rangle$ gives the expected signal energy on top of node $j$ and the $jk$ off-diagonal element gives $\langle j| \hat{\mathbf{U}}_\tau |k\rangle=\sum\limits_{i}p_{i}\langle j| \hat{\mathbf{U}}^{(i)}_\tau |k\rangle$ gives the expected covariance between nodes $j$ and $k$. From the statistical propagator, it is straightforward to obtain the density matrix. Here, the trace of the statistical propagator plays the role of the partition function giving the total expected signal energy in the system $Z_{\tau} = \tr{\hat{\mathbf{U}}_{\tau}}=\sum\limits_{i,j}p_{i}\langle j| \hat{\mathbf{U}}^{(i)}_\tau |j\rangle$ and can be used to normalize the statistical propagator, fixing the expected signal energy in the system to be one unit. Therefore, the density matrix gives the statistics of system---e.g., expected covariance between nodes, signal distribution and average response diversity--- for one unit of signal energy (See Fig.~\ref{fig:static_networks_density})

\begin{equation}
    \hat{\boldsymbol{\rho}}_{\tau} = \frac{\hat{\mathbf{U}}_{\tau}}{Z_{\tau}}.
\end{equation}

In contrast with the statistical propagator, the diagonal elements of the density matrix admit a probabilistic interpretation. For instance, the $i$-th diagonal element $\langle i|\hat{\boldsymbol{\rho}}_\tau|i\rangle$ is the probability of finding one unit of signal energy on top of node $i$. It is important to note that with our approach, based on summation of outer product of vectors with themselves, this formulation of density matrix is always positive semi-definite, regardless of the dynamics and the features of the underlying network. Furthermore, the weighted summation of local propagators gives high explanatory power to the density matrix, describing the statistics of systems' response to stochastic perturbations happening at different sites. The diversity of response to perturbations (See Fig.~\ref{fig:schematic}) which, similarly to previous works~\cite{ghavasieh2020statistical,ghavasieh_density_brain,ghavasieh_fungal}, can be quantified in terms of the Von Neumann entropy $S_\tau = - \tr{\hat{\boldsymbol{\rho}}_\tau\log{\hat{\boldsymbol{\rho}}_\tau}}$ tends to be high when the perturbed states are distinct and diverse and the surprise--- i.e., number of bits required to describe the system's response to perturbations quantified by the Von Neumann entropy--- in observing a response to perturbation is high (See Fig.~\ref{fig:schematic}). 

Note that in our formulation, the field is not required to be positive and real, and it can have negative or imaginary values. Yet, remarkably, this approach is in agreement with the original formulation, if the control operator is Hermitian $\hat{\mathbf{H}}=\hat{\mathbf{H}}^{\dag}$(See Appendices). We report control operators for: i) a number of linear dynamical processes, including diffusion, discrete and continuous random walks, graph walks and consensus dynamics (See Tab.~\ref{tab:Linear_Dynamics}) and ii) a number of dynamics linearized close to steady state, including biochemical, birth-death, regulatory, epidemics, synchronization, mutualistic, neuronal and voter models (See Tab.~\ref{tab:NONLinear_dynamics}). In general, assuming that the probability of perturbing the nodes is uniform $p_i=1/N$, and the dynamical equation is continuous and linear (or linearized) $\partial_\tau|\psi_{\tau}\rangle =-\hat{\mathbf{H}}|\psi_{\tau}\rangle $, the statistical propagator takes the form of $\hat{\mathbf{U}}_\tau = \hat{\mathcal{G}}_\tau \hat{\mathcal{G}}_{\tau}^{\dag} = e^{-\tau\hat{\mathbf{H}}}e^{-\tau\hat{\mathbf{H}}^{\dag}}$. In case of linear discrete dynamics, like $|\psi_{\tau+1}\rangle =\hat{\mathbf{H}}|\psi_{\tau}\rangle$, with the discrete time-evolution operator $\hat{\mathcal{G}}_\tau = \hat{\mathbf{H}}^{\tau}$, with $\tau$ taking only non-negative integers and the control operator $\hat{\mathbf{H}}$ being a transition matrix, the statistical propagator reads $\hat{\mathbf{U}}_\tau = \hat{\mathcal{G}}_\tau \hat{\mathcal{G}}_{\tau}^{\dag}=\hat{\mathbf{H}}^{\tau}\hat{\mathbf{H}}^{\tau\dag}$ (For more details, see Appendices).

\begin{figure*}
\centering
\includegraphics[width=.7\textwidth]{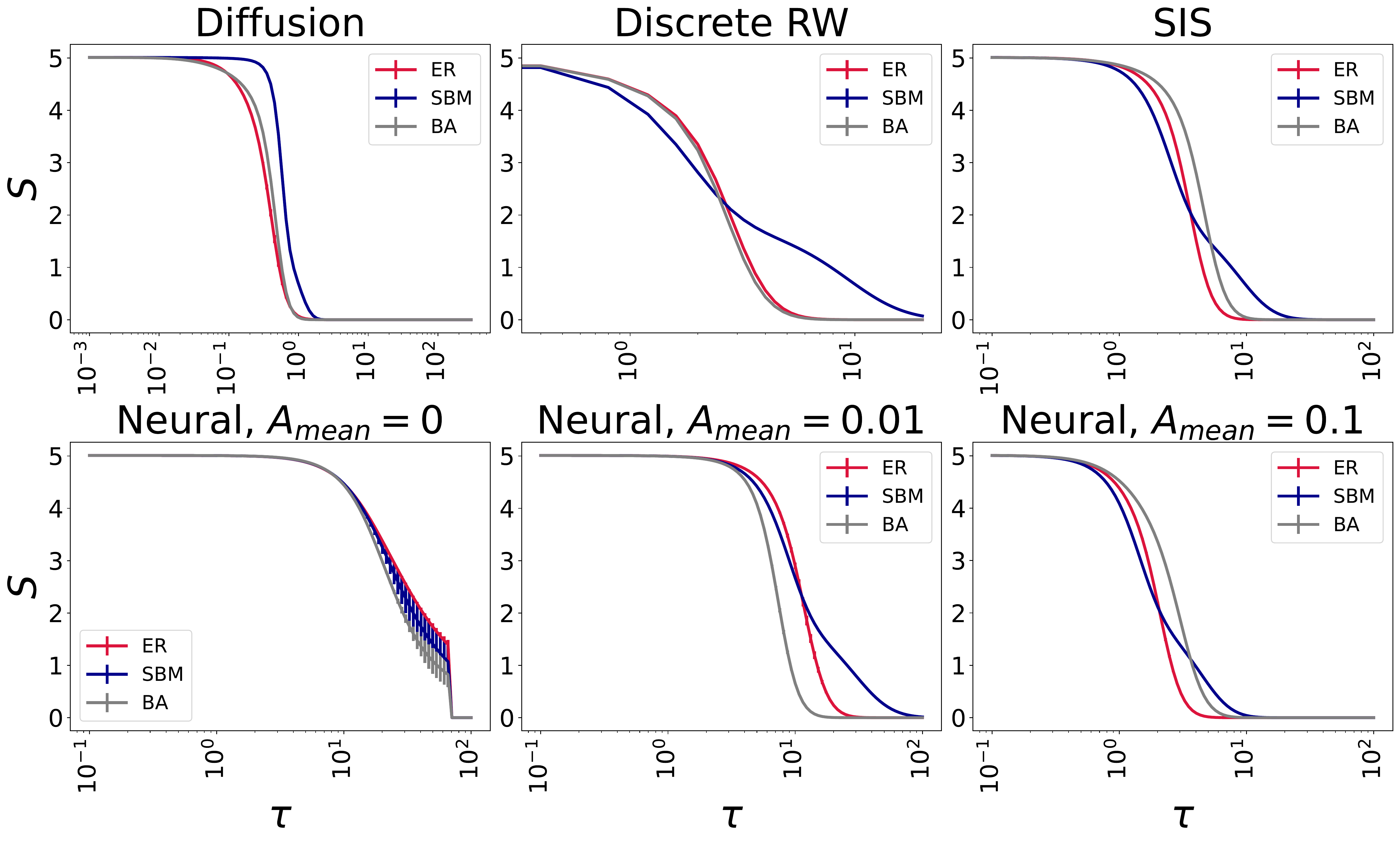}
\caption{\label{fig:static_networks_entropy}\textbf{Response diversity to perturbations.} We compare the functional diversity of different systems, with structures given by Erdos-Renyi (ER), Barabasi-Albert (BA) and stochastic block model (SBM) and different dynamics: continuous diffusion, discrete random walk (RW), SIS epidemics and neural. For neural dynamics, we assigned random negative and positive weights to links, drawn from a Gaussian distribution of mean values of $0, 0.01, 0.1$ and standard deviation of $1/N$. Generally, SBM has an advantage in keeping the functional diversity at larger propagation scales $\tau$. Yet, at smaller propagation scales, the best topological feature highly depends on the dynamics. For instance, in case of SIS dynamics, it seems that degree heterogeneity plays an important role at small scale propagation. Conversely, the homogeneous distribution of links in ER provides a more diverse response to perturbation, if neural dynamics is considered and the network is dominated by excitatory links--- i.e., most weights are positive. This figure shows that by changing the propagation scale $\tau$ or the dynamical process, each topological feature--- e.g., randomness, modularity, heterogeneity, presence of negative links, asymmetries--- can become advantageous or disadvantageous for keeping the information flow diverse.}
\end{figure*}

\section{Synthetic network analysis.} Here, we firstly study three different dynamics, that could not be considered within the previous perspective, including discrete random walks (RW), Susceptible-Infectious-Susceptible (SIS) epidemic spreading and the large scale neuronal dynamics of brain regions (See Tab.~\ref{tab:Linear_Dynamics} and Tab.~\ref{tab:NONLinear_dynamics}
). The set of parameters we choose for the neuronal dynamics is $\{B = 3, C = 1, R = 2.5\}$ with an initial state drawn from a Gaussian distribution with $x_{mean}=0$ and $x_{std}=1/N$. Similarly, the SIS parameters are given by $\{B=1, R=1/10 \}$ to keep it bellow the critical threshold of spreading, and the initial state is also drawn from a Gaussian distribution with $x_{mean}=1/10$ and $x_{std}=1/N$. The dynamics is run on top of three different types of static networks with $N=150$, including an Erdos-Renyi (ER) network with connectivity probability of $0.05$, a stochastic block model (SBM) with five modules where the intera-community probability of connection is respectively $0.015$ and the inter-community probability is $0.4$, and a Barabasi-Albert (BA) network with $m=6$. The choice of parameters keeps the average degree $\langle k \rangle \approx 7.8$, for all network types and realizations. Note that for neural dynamics, in accordance with other studies~\cite{Neural_bistable}, we randomly assign positive and negative weights to the edges of the network drawn from a Guassian distribution with $A_{mean}=0$ and $A_{std}=10/N$, that allows for having both excitatory and inhibitory interactions between the nodes and making the underlying network nonsymmetric (directed). Similarly, we analyze other choices of mean value ($A_{mean}=0,0.01,0.1$), later, in the analysis of functional diversity. To find the linear response to perturbation in neural dynamics, we assume a random initial condition taken from Gaussian distributions with meanvalue of $0$ and standard deviation of $1/N$ and calculate its steady state for each network. Similarly, for SIS epidemics, we assume a random initial condition taken from Gaussian distributions with meanvalue of $0.1$ and standard deviation of $1/N$ and calculate its steady state for each network. For both cases, we calculate the Jacobian matrix $-\hat{\mathbf{H}}$ (See Appendices) according to Tab.~\ref{tab:NONLinear_dynamics} and calculate the statistical propagator with $p_i=1/N$. For random walk dynamics, as the master equation is linear, there is no need for linearization. Consequently, we derive the density matrices for all $3$ dynamics on top of the $3$ synthetic networks (See Fig.~\ref{fig:static_networks_density}). Also, to study their functional diversity, we calculate the Von Neumann entropy for $10$ realizations of each of these three cases and report the average value and the variance (See Fig.~\ref{fig:static_networks_entropy}). 

\begin{figure*}
\centering
\includegraphics[width=\textwidth]{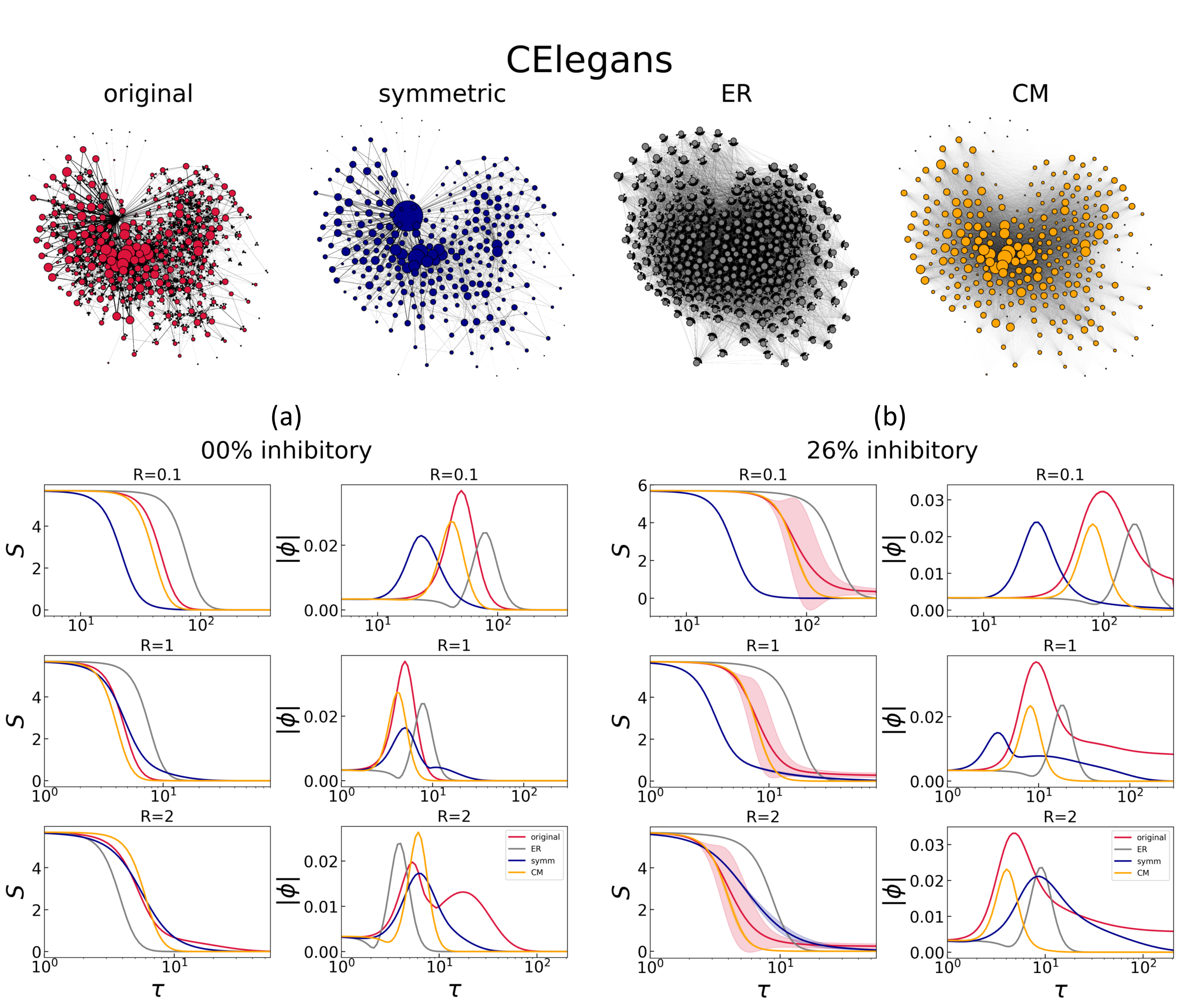}
\caption{\label{fig:CElegans}\textbf{Functional diversity of C. Elegans neuronal network} On the top, networks representing the structure of the C. Elegans nervous system (original), with its corresponding null models ER, CM and symmetric (for details, see text) are represented. (a) Shows the functional diversity $S$ and the fragility of functional diversity $|\phi|$ at different propagation scales $\tau$ for all four networks, for different values of $R=0.1,1,2$. (b) Shows the same, but after turning $26\%$ of overall weights negative (inhibitory). The lines and shades show, respectively, mean values and variances, over $20$ realizations. Interestingly, in absence of inhibitory connections, ER shows the highest and lowest functional diversity depending on the choice of the interaction coefficient $R$. Also, in this regime, the symmetric null model has a functional diversity comparable to the original networks, only if the interaction coefficient $R$ is not small. Instead, for small interaction coefficients, CM generates the information dynamics most similar to the original network, at all propagation scales. In presence of inhibitory connections, we observe a totally different behavior for the original network and its null models. Also, the fragility of functional diversity, an important indicator of how system can stay operative in risky environment, has global maxima at different propagation scales for different networks with distinct dynamical coefficients, which is higher for the original network compared to the others in almost all cases considered.}
\end{figure*}

\begin{figure*}
\centering
\includegraphics[width=\textwidth]{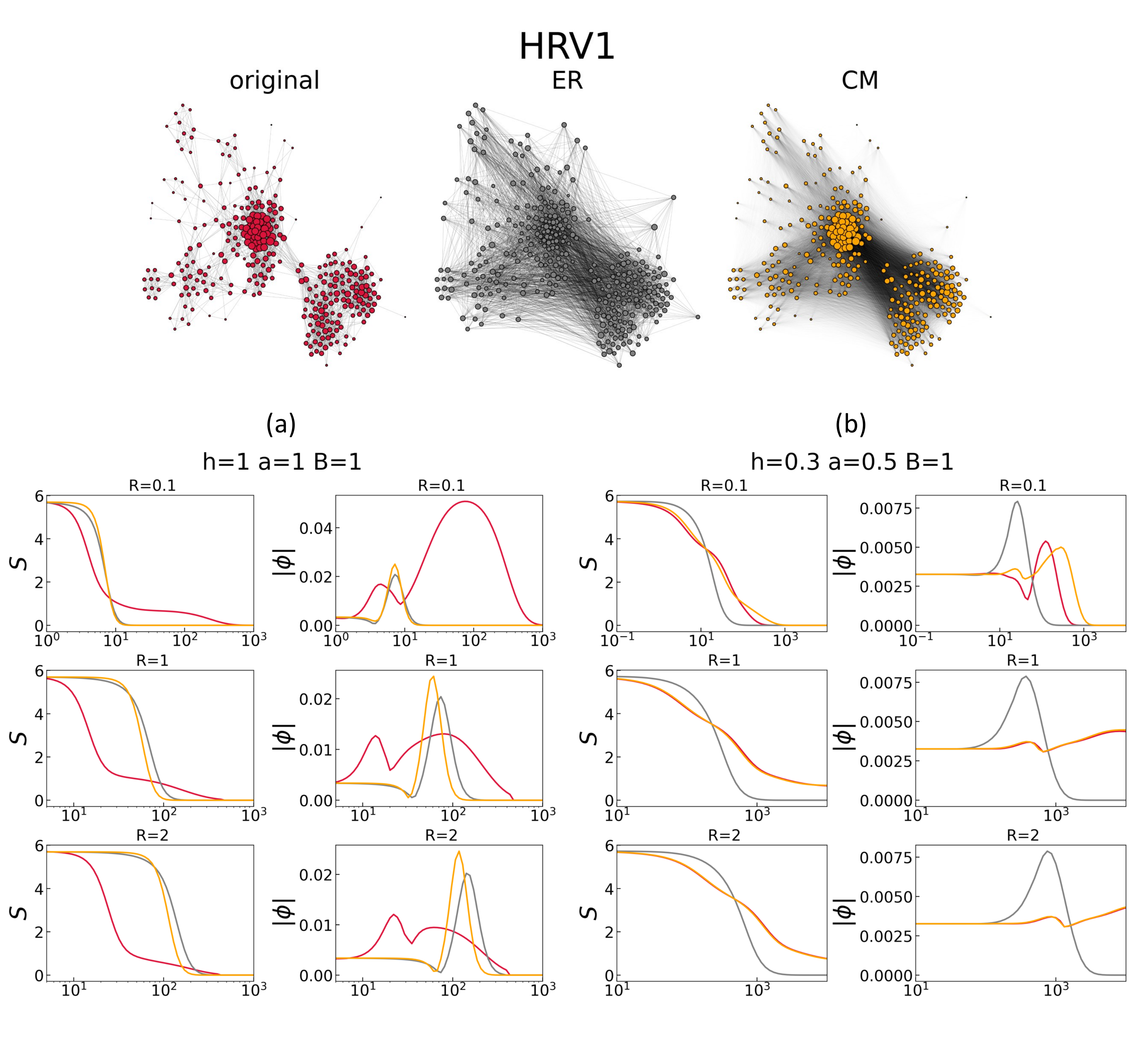}
\caption{\label{fig:HRV1}\textbf{Functional diversity of HRV1 gene regulatory network.} On the top, networks representing the structure of the HRV1 nervous system (original), with its corresponding null models ER and CM (for details, see text) are represented. (a) Shows the functional diversity $S$ and the fragility of functional diversity $|\phi|$ at different propagation scales $\tau$ for all three networks, for different values of $R=0.1,1,2$ and fixed values $h=1,a=1,B=1$ in the dynamical equation (See Tab.\ref{tab:NONLinear_dynamics}). (b) Shows the same, but for different fixed values $h=1/3,a=1/2,B=1$. The lines and shades show, respectively, mean values and variances, over $20$ realizations. While for the parameters of column (b) the configuration model (CM) provides a good model, it fails with other parameters ( column a) where it behaves more like ER and the two behave radically different from the original network. Column (a), also, shows an interesting behavior for the fragility of functional diversity: In contrast with the null models, there are two peaks for the original network and the decay of fragility happens slowly.}
\end{figure*}

\section{Biological systems.} We consider the structure of the nervous system of the nematode Caenorhabditis elegans~\cite{CElegans_data,Watts1998} with $N=297$ nodes and average degree of $\langle k \rangle = 7.94$. This network is weighted, with weights indicating the number of connections between neurons, and directed, meaning that some of the connections can be from one neuron to another in an asymmetric way. To understand the effect of different topological features of the neural system on its functional diversity, we compare it with a number of null models. The "ER" null model contains the same total weight, but each link with unit weight, is connecting a pair of randomly chosen nodes. This null model destroys almost all complex features, most notably the heterogeneity of degree distribution. Another null model of interest is the configuration model, "CM", where the degree (strength) distribution is kept, but the connections are randomized. It is straight forward to obtain the adjacency of CM from the adjacency of the original network $A$, as: $k_i k_j/2m$, where $k_i$ is the strength of node $i$ and $m$ is the total strength of links. Finally, to study the effect of directed connections on the functional diversity, we generate the symmetric model, whose adjacency is simply $(A+A^{\dag})/2$, keeping the total strength and degree heterogeneity, while symmetrizing the asymmetric connections. 

Since, in contrast with macroscopic networks of brain areas,  there is no local self-excitation in microscopic neural networks like C. Elegans connectome, we remove the self-excitation term ($C=0$) in the equation governing neuronal dynamics (See Tab.~\ref{tab:NONLinear_dynamics}) similar to one of the references~\cite{Neural_bistable}. Also, we fix one of the parameters $B=1$ that encodes the self-inhibition of neurons, and study the dynamics for different values of the interaction parameter $R=0.1,1,2$. For each network, we give a random initial condition taken from Gaussian distributions with meanvalue of $0.01$ and standard deviation of $1/N$ and calculate its steady state for each network. From here, it is straightforward to calculate the Jacobian, which is the negative of the control operator operators $\hat{\mathbf{H}}$, required to calculate the functional diversity in terms of the Von Neumann entropy $S_\tau$. In addition to the functional diversity, we calculate the fragility of functional diversity, which is the average change in the Von Neumann entropy due to node removals. For instance, assume node $i$ is removed from the network and the Von Neumann entropy after removal of node $i$ changes by $\delta S_{\tau}^{(i)}$. It is worth noting that $\delta S^{(i)}_{\tau}$ has been previously used as a multiscale centrality measure to devise attack strategies, highly effective in dismantling structures and information dynamics, outperforming state-of-the-art methods~\cite{ghavasieh_Structural_robustness,ghavasieh_functional_robustness}. The average fragility of functional diversity is given by $|\phi_\tau|=\frac{1}{N}\sum\limits_{i=1}^{N}|\delta S^{(i)}_{\tau}|$. We run $20$ realizations of the above procedure and report the average and variance of $S_\tau$ and $|\phi_\tau|$ in Fig.~\ref{fig:CElegans}. Furthermore, the studies suggest that around $26\%$ of links are inhibitory in the C. Elegance network~\cite{26CElegans}. Fortunately, the framework is capable with signed network including negative and positive links, as shown in previous sections. Therefore, we randomly select $26\%$ of the overall weights in the original and null model networks, and multiply them by $-1$ to turn them inhibitory and make an ensemble of $20$ realizations for each network. The average and variance of functional diversity and its fragility can be found in Fig.~\ref{fig:CElegans}. 

Similarly, we analyse a network of malaria parasite genes (HRV1)~\cite{malaria_netwrok} with $N=307$ nodes and average degree of $\langle k \rangle$. Since the network is undirected, the null model networks against which we compare are limited to ER and CM, generated as for C. Elegans. We run a Gene-Regulatory dynamics on top of these network (See Tab.~\ref{tab:NONLinear_dynamics}) with initial conditions drawn from Gaussian distribution of mean $0.2$ and standard deviation of $1/N$ and a range of parameters (See Fig.~\ref{fig:HRV1}), adopted from other studies~\cite{Barzel_universality,Barzel_minimal}. 

\section{Discussion.} 

The original formulation of the network density matrix~\cite{de2016spectral,ghavasieh2020statistical,Ghavasieh_perspective} has been successful in describing the dynamics of information on top of complex structures, enabling advance structural and functional robustness~\cite{ghavasieh_Structural_robustness,ghavasieh_functional_robustness} and reducibility~\cite{de2015reducibility,ghavasieh2020enhancing} analysis, (dis)similarity assessment~\cite{de2016spectral,ghavasieh2021multiscale,ghavasieh_density_brain}, characterization of criticality in networks~\cite{Gabrielli_phase_transition} with potential for machine learning methods. However, here we provide a detailed discussion of the shortcomings of this framework, mainly its limitation to study dynamics other than diffusion and structures with asymmetries. To take another step towards a better understanding of how complex systems operate, in terms of density matrices, we use information theory and signal processing to extend the method's applicability to a much wider range of dynamics and classes of structure, including directed and signed ones. 

This new framework considers stochastic perturbations at different locations in the system, propagating according to linear, linearized or non-linear dynamical laws, constrained by structural links and paths. The propagation of perturbations from each node $i$ proxies information flow from that node to others and can be encoded in a local propagator matrix, where diagonal elements give the signal energy and the off-diagonal elements are covariance between pairs of nodes in receiving signal amplitudes from $i$. Considering the stochastic nature of complex systems, we consider that propagations start from different localities indeterminately, with given probabilities. Finally, we obtain the density matrix as a normalized superposition of propagation patterns, each weighted their occurrence probability. Such density matrix can have tremendous power in describing the statistics of systems' response to stochastic perturbations, in terms of signal energy distribution, node-node correlation and heterogeneity of response to stimuli. An interesting byproduct it that our framework is compatible with the standard quantum density matrix approach, thus providing an opportunity to explore the bridge between complexity science and quantum statistical physics (See Appendices), opening the doors for future cross pollination. A schematic illustration of the network density matrix idea is presented in Fig.~\ref{fig:schematic} and, for instance, density matrices of three classes of synthetic networks, including random, modular and heterogeneous coupled with random walks, SIS epidemics and neural dynamics in presence of excitatory and inhibitory connections is are shown in Fig.~\ref{fig:static_networks_density}.

To show how the framework can be used for practical purposes, across distinct systems, we study the functional diversity---i.e., complexity and heterogeneity of response to internal and external stimuli or perturbations---, a prominent feature of complex systems. For instance, in a biological network like the human connectome, the extraordinary range of physiological response enables the system to generate and distribute information and coordinate its activity at different scales~\cite{discovering_connectome}. Functional diversity can be measured in bits by means of the Von Neumann entropy of the statistical ensemble~\cite{Ghavasieh_perspective}--- as it directly measures the average information--- i.e., log probability--- in observing patterns of activity in response to perturbations. Therefore, as a direct application of this framework, we analyze the functional diversity of the mentioned syntethic networks coupled with a range of dynamics, including continuous diffusion, discrete random walks, SIS epidemics and neural dynamics in absence and presence of inhibitory connections. Interestingly, according to our results, it is not trivial to tell in advance which class of network has the highest functional diversity (See Fig.~\ref{fig:static_networks_entropy}). While in most cases the modular structure provides the best solution for long range signal propagation, at the small or middle propagation scales, randomness, modularity and heterogeneity compete, closely. The result clearly shows that the diversity of response to perturbation can not be determined by topological features nor the dynamical rules, but their coupling. In other words, by changing the propagation scale $\tau$, or dynamical processes, one observes that different topological features--- e.g., randomness, modularity, heterogeneity, presence of negative links, asymmetries--- become relevant or irrelevant for maintaining the functional diversity.

Similarly, we analyze the propagation of stochastic perturbations in empirical biological networks including the nervous system of C. Elegans and the gene regulatory network HRV1. In addition to the functional diversity, we study the fragility of functional diversity under structural damage--- i.e., random node removals. In both cases, we compare the empirical networks with weak and strong null models capturing distinct topological features. We show that the success of each null model in estimating the system statistics strongly depends on the parameters of the dynamical equation and the propagation scale $\tau$ (See Fig.~\ref{fig:CElegans}). For instance, in absence of inhibitory connections in C. Elegans, a random null model (ER) shows the highest and lowest functional diversity depending on the choice of the interaction coefficient $R$ in the neural dynamics equation. Moreover, the symmetric null model which is identical to the original C. Elegans network but with a symmetrized adjacency matrix, provides a good model for the original networks, only if the interaction coefficient $R$ is large. Instead, for small interaction coefficients, the Von Neumann entropy of the configuration model, the null model that only preserves the degree heterogeneity and randomizes all other features, behaves more similarly to the original network, at all propagation scales. In presence of $26\%$ inhibitory connections, as estimated for C. Elegans connectome~\cite{26CElegans}, the behavior of null models and the original network completely changes. Also, the fragility of functional diversity, an important indicator of how system can stay operative in unsafe environment, has global maxima at different propagation scales for different networks with different dynamical coefficients. The analysis of the HRV1 gene regulatory network confirms the previous results (See Fig.~\ref{fig:HRV1}). While for certain parameters the configuration model (CM) provides a better estimation than the random null model, it fails in other scenarios where the two behave radically different. Yet, in this case, similar to the synthetic network analysis provided before (See Fig.~\ref{fig:static_networks_entropy}), it seems that the modular organization of the structure of the original network keeps the functional diversity high at large propagation scales.

Our analysis of synthetic and empirical biophysical networks clearly shows that deducing the functional diversity directly from the mere topological features or dynamical rules on their own is impossible. Instead, the network density matrix formalism, now extended to include a broad and rich range of dynamics and structural categories, provides a versatile approach to tackle problems dealing with the complex interplay between structure and dynamics, shedding light on how complex systems operate and suggesting that structural information alone is not sufficient to characterize empirical systems.

\appendix

\section{Review of the original formulation of network density matrices.} In this subsection, we briefly review the fundamentals of the original framework~\cite{ghavasieh2020statistical}. We encode the nodes as orthogonal canonical vectors $| i\rangle, (i= 1,2,...N$) with $\langle i|j\rangle =\delta_{ij}$ being the delta function, equal to 0 for $i\neq j$ and 1 only if $i=j$. A field $| \phi_\tau\rangle$ is assumed to be on top of the network with concentration of $\langle i |{\phi}_\tau\rangle$ on top of $i$-th node at time $\tau$. The evolution of the field is assumed to be governed by
\begin{equation}\label{eq:master_original}
\partial_{\tau}|\phi_\tau\rangle=-\hat{\mathbf{H}}|\phi_\tau\rangle,
\end{equation}
with $\hat{\mathbf{H}}$ being the control operator. Solving Eq.~\ref{eq:master_original} one finds the time-evolution operator of the dynamics

\begin{equation}\label{eq:propagator_original}
\hat{\mathcal{G}}_{\tau} = e^{-\tau \hat{\mathbf{H}}},
\end{equation}
whose $ij$--th element encodes the flow of field from node $i$ to node $j$, also written as $\langle j|\hat{\mathcal{G}}_{\tau}|i\rangle$.
From the time evolution operator, it is straightforward to derive the density matrix 
\begin{equation}\label{eq:density_original}
\hat{\boldsymbol{\rho}}_{\tau} = \frac{\hat{\mathcal{G}}_{\tau}}{\tr{\hat{\mathcal{G}}_{\tau}}},
\end{equation}
and the normalization factor $Z_\tau = \tr{\hat{\mathcal{G}}_{\tau}}$ is the partition function of the system encoding the dynamical trapping--- i.e., the amount of field that is still on top of the initiator node. It is worth remarking that this operator is formally equivalent to the network density matrix introduced in Ref.~\cite{de2016spectral}, in special case where $\hat{\mathbf{H}}$ is the combinatorial Laplacian matrix. The Von Neumann entropy of the density matrix is given by

\begin{equation}\label{eq:entropy_original}
S_{\tau} = -\tr{ \hat{\boldsymbol{\rho}}_{\tau} \log{\hat{\boldsymbol{\rho}}_{\tau}}   }.
\end{equation}

Finally, a diagonalizable time-evolution operator can be eigen decomposed as $\hat{\mathcal{G}}_\tau= \sum\limits\limits_{\ell=1}^{N} \alpha_{\ell}(\tau)\hat{\boldsymbol{\sigma}}^{(\ell)}$, where $\alpha_{\ell}(\tau)$ is the $\ell$-th eigenvalue of the time-evolution operator, and $\hat{\boldsymbol{\sigma}}^{(\ell)}$ is the outer product of its $\ell$-th right and left eigenvectors of $\hat{\mathbf{H}}$. Similarly, the eignevalues of the density matrix are given by $\rho_\ell(\tau) = \alpha_{\ell}(\tau)/Z_{\tau}$. Each operator $\hat{\boldsymbol{\sigma}}^{(\ell)}$ obtained from eigen-decomposition of the time-evolution operator act like a stream, guiding the flow, and is multiplied by its contribution $\alpha_{\ell}(\tau)$. In Ref.~\cite{ghavasieh2020statistical} the authors start from Eq.~\ref{eq:master_original} to introduce the \emph{stream operators} defined as $\hat{\boldsymbol{\sigma}}^{(\ell)}\in \mathbb{R}^{N\times N}$ which guide the propagation of perturbations of the field on the top of the network. Since each stream operator $\ell$ has activation probability $\rho_{\ell}(\tau)\in\mathbb{R}$, they work as a statistical ensemble whose superposition shapes the density matrix
\begin{eqnarray}
\hat{\boldsymbol{\rho}}_\tau = \sum\limits_{\ell=1}^{N}\rho_{\ell}(\tau)\hat{\boldsymbol{\sigma}}^{(\ell)}.
\end{eqnarray}

\section{Statistical ensemble microstates.} Here, we provide an alternative way to interpret the probabilities, in terms of microstates. As explained in the previous subsection, a diagonalizable time-evolution operator Eq.~\ref{eq:propagator_original} is the solution of Eq.~\ref{eq:master_original}, describing the evolution of the field
\begin{eqnarray}
\ket{\phi_\tau}=\hat{\mathcal{G}}_\tau\ket{\phi_{0}}=\sum\limits_{\ell=1}^{N}\alpha_{\ell}(\tau)\hat{\boldsymbol{\sigma}}^{(\ell)}\ket{\phi_{0}}.
\end{eqnarray}

Note that $\hat{\boldsymbol{\sigma}}^{(\ell)}\ket{\phi(0)}=\ket{v_{\ell}}\braket{v_{\ell}}{\phi(0)}$, where $\ket{v_{\ell}}$ is an eigenstate of $\hat{\mathbf{H}}$ and the second factor $\braket{v_{\ell}}{\phi(0)}$ is the projection of the initial field $\ket{\phi_{0}}$ along the the $\ell$--th eigenstate of $\hat{\mathbf{H}}$. This fact leads to the following result: the field configuration at time $\tau$ is the evolution of the superposition of its initial configurations projected along the $\hat{\mathbf{H}}$ eigenstates.

In quantum physics, the microscopic variables of the system are the amplitudes of the basis function. For instance, if the basis is chosen to be the eigenstates of position, then the microscopic variables are the values of the wave function in each point in space. Here, we can use a similar argument to identify the microscopic variables in the SFT as the projections of $\ket{\phi_\tau}$ onto a basis. If we choose $\{\ket{v_{\ell}}\}$ for this purpose, then the microscopic variables are given by
\begin{eqnarray}
\braket{v_{\ell}}{\phi_\tau}&=&\bra{v_{\ell}}\sum\limits_{\ell'=1}^{N}\alpha_{\ell}(\tau) \ket{v_{\ell'}}\braket{v_{\ell'}}{\phi_{0}}\nonumber\\
&=&\sum\limits_{\ell'=1}^{N}\alpha_{\ell}(\tau) \braket{v_{\ell}}{v_{\ell'}}\braket{v_{\ell'}}{\phi_0}\nonumber\\
&=& \alpha_{\ell}(\tau)\braket{v_{\ell}}{\phi_0}.
\end{eqnarray}
It follows that the \emph{microstates}, in our framework, are the $N$ amplitudes obtained by projecting the initial field configuration onto the $\hat{\mathbf{H}}$ eigenstates and evolved until time $t$ by the corresponding time-evolution operator. 

The number of such microstates can be effectively quantified, at time $\tau$, as $\sum\limits\limits_{\ell=1}^{N}\alpha_{\ell}(\tau)$ which, in fact, corresponds to the partition function $Z_\tau$ introduced and described in Refs.~\cite{de2016spectral,ghavasieh2020statistical,de2019quantum_networks,Ghavasieh_perspective}. Therefore, the probability of each microstate is given by 
\begin{eqnarray}
\rho_{\ell}(t) = \frac{\alpha_{\ell}(\tau)}{\sum\limits\limits_{\ell'}\alpha_{\ell'}(\tau)}.
\end{eqnarray}
Interestingly, this result opens the door to the possibility of identifying $\hat{\boldsymbol{\sigma}}^{(\ell)}$ with a microstate, and to define the state of the system in terms of the superposition of all microstates when it is not possible to know the current microstate before performing a measure. In practice, our best choice to describe the system is to estimate the expected state as the weighted average
\begin{eqnarray}
\hat{\boldsymbol{\rho}}(\tau) =\sum\limits_{\ell=1}^{N}\rho_{\ell}(\tau) \hat{\boldsymbol{\sigma}}^{(\ell)}.
\end{eqnarray}
This last operator, which encode the expected state of the system obtained from the ensemble of microstates, is formally equivalent to the density matrix used in quantum physics, as well as to the network density matrix introduced in~\cite{de2016spectral} and later understood in terms of a statistical field theory in~\cite{ghavasieh2020statistical}.

\section{Case of Hermitian control operator.} For hermitian $\hat{\mathbf{H}}$, the eigenvalues of the time-evolution operator are real and non-negative. An example can be the diffusion dynamics on top of directed static networks. In this case, the partition function and Von Neumann entropy are real and non-negative, being summations of non-negative values, and the eigenvalues of the density matrix can be interpreted as probabilities of activation of streams (See ... for a complete description), providing a valid statistical description of the information dynamics. 

\section{Case of non-Hermitian control operator.} Finding the statistics of non-Hermitian Hamiltonians is a hot topic in quantum physics. Similarly, a number of important linear dynamics are governed by non-hermitian control operators $\hat{\mathbf{H}}$, like diffusion on top of directed networks. It is important to note that in these cases the eigenvalues of density matrix can be imaginary and, therefore, it would be difficult to interpret them as activation probabilities for streams. However, here we show that the Von Neumann entropy and parition functions of such systems are real.

Assume the elements of the time-evolution operator are all real values. Thus, the complex eigenvalues of the time-evolution operator are known to be complex conjugate pairs. For instance, assume the $\ell$-th eigenvalue is given by $\alpha_{\ell}(\tau)= \alpha_{\ell}^{Re}(\tau) + i \alpha_{\ell}^{Im}(\tau) = r_\ell(\tau) e^{i \theta(\tau)}  $, where $\alpha_{\ell}^{Re}(\tau)$ and $\alpha_{\ell}^{Im}(\tau)$ are, respectively, the real and imaginary parts of the eigenvalue and $r_\ell(\tau)$ and $e^{i \theta(\tau)} $ provide the polar representation. Since the complex eigenvalues are complex conjugate pairs, there must be an eigenvalue $\alpha_{\ell'}(\tau)= \alpha_{\ell}^{Re}(\tau) - i \alpha_{\ell}^{Im}(\tau) = r_{\ell}(\tau) e^{-i \theta(\tau)}  $. It is straightforward to show that the summation of these two eigenvalues is the summation of the real parts $\alpha_{\ell}(\tau)+\alpha_{\ell'}(\tau)=2\alpha_{\ell}^{Re}(\tau)$. Therefore, the partition function can be written as
\begin{equation}
Z_{\tau} = \tr{\hat{\mathcal{G}}_\tau} = \sum\limits_{\ell}\alpha_{\ell}^{Re}(\tau),
\end{equation}
which is always a real number.

The logarithm of the eigenvalues of the density matrix, used to calculate the Von Neumann entropy, can be written as $\log{r_{\ell}e^{i \theta_{\ell}}/Z_\tau}=\log{r_{\ell}}+i \theta_{\ell}-\log{Z_\tau}$. For the complex conjugate eigenvalue, it is given by $\log{r_{\ell}e^{-i \theta_{\ell}}/Z_\tau}=\log{r_{\ell}}-i \theta_{\ell}-\log{Z_\tau}$. The summation of these two multiplied by the corresponding eigenvalue of the time-evolution operator matrix
leads to

\begin{eqnarray}
\alpha_{\ell}(\tau)\log\alpha_{\ell}(\tau)+\text{c.c.}&=&2\alpha_{\ell}^{Re}(\tau) \log{r_\ell} - 2 \alpha_{\ell}^{Re}(\tau) \log{Z_\tau} \nonumber\\
&-& 2 \alpha_{\ell}^{Im}(\tau)\theta_{\ell}(\tau)\nonumber
\end{eqnarray}
where all the imaginary parts cancel out.,
Therefore, the Von Neumann entropy is real and, yet, depending on the eigenvalues, it is possible to obtain a negative value which is cannot be trivially interpreted, at the moment, in probabilistic terms. In the following, we will better understand under which conditions non-Hermitian control operators will still lead to non-negative entropy.

\section{Jordan decomposition for the general case.} An even more general result can be obtained by considering that the control operator $\hat{\mathbf{H}}'$ is a generic matrix that can be decomposed into its Jordan normal form over the field of complex numbers as $\hat{\mathbf{H}}'=\hat{\mathbf{P}}^{-1}\hat{\mathbf{J}}\hat{\mathbf{P}}$, where $\hat{\mathbf{J}}=\text{diag}(\hat{\mathbf{J}}_{1},\hat{\mathbf{J}}_{2},...,\hat{\mathbf{J}}_{k})$ is a block matrix and $\hat{\mathbf{J}}_{k}$ is the matrix whose principal diagonal entries are equal to $\alpha_k$ and the upper diagonal entries are equal to $1$. For sake of simplicity, let us consider that $\hat{\mathbf{H}}'=-\tau \hat{\mathbf{H}}$, leading to a Jordan block matrix $\hat{\mathbf{J}}(\tau)$. It follows that
\begin{eqnarray}
e^{\hat{\mathbf{H}}'}=\hat{\mathbf{P}}^{-1}e^{\hat{\mathbf{J}}(\tau)}\hat{\mathbf{P}}\nonumber
\end{eqnarray}
leading to entropy
\begin{eqnarray}
S_{\tau}=\log Z_{\tau} - Z_{\tau}^{-1}\tr{e^{\hat{\mathbf{J}}(\tau)} \hat{\mathbf{J}}(\tau)}.
\end{eqnarray}
Since
\begin{eqnarray}
e^{\hat{\mathbf{J}}_{k}(\tau)}=e^{\alpha_{k}(\tau)}\left(
\begin{matrix}
\frac{1}{0!} & \frac{1}{1!} & ... & \frac{1}{(\ell-1)!} \\
0 & \frac{1}{0!} & ... & \vdots \\
\vdots & \ddots & \frac{1}{0!} & \frac{1}{1!} \\
0 & ... & ... & \frac{1}{0!} 
\end{matrix}\right)\nonumber
\end{eqnarray}
and
\begin{eqnarray}
\hat{\mathbf{J}}_{k}(\tau)=\alpha_{k}(\tau)\left(
\begin{matrix}
1 & \frac{1}{\alpha_{k}(\tau)} & ... & 0 \\
0 & 1 & ... & \vdots \\
\vdots & \ddots & 1 & \frac{1}{\alpha_{k}(\tau)} \\
0 & ... & ... & 1 
\end{matrix}\right),\nonumber
\end{eqnarray}
it follows that 
\begin{eqnarray}
\tr{e^{\hat{\mathbf{J}}(\tau)} \hat{\mathbf{J}}(\tau)}&=&\sum_{k\in\text{blocks}}\tr{e^{\hat{\mathbf{J}}_{k}(\tau)} \hat{\mathbf{J}}_{k}(\tau)}\nonumber\\
&=&\sum_{k\in\text{blocks}}\mu_{k}e^{\alpha_{k}(\tau)} \alpha_{k}(\tau),\nonumber
\end{eqnarray}
where $\mu_{k}$ is the rank of the $k$--th block. Note that if there are $N$ rank-1 blocks, corresponding to the case of an Hermitian operator, then the result will coincide with the expected one~\cite{de2016spectral,ghavasieh2020statistical}.

Let us focus at a higher level of detail as follows. The partition function is given by
\begin{eqnarray}
\label{eq:Z-Jordan-form}Z_{\tau}=\sum_{k\in\text{blocks}}\mu_{k}e^{\alpha_{k}(\tau)},
\end{eqnarray}
where we expect complex conjugate pairs to be eigenvalues of the control operator. Since $\alpha_{k}(\tau)=\alpha_{k}^{Re}(\tau)+\alpha_{k}^{Im}(\tau)$, we can write
\begin{eqnarray}
Z_{\tau}=\sum_{k\in\text{blocks,Re}}\mu_{k}e^{\alpha_{k}(\tau)} + \sum_{k\in\text{blocks,Im}}\mu_{k}(e^{\alpha_{k}(\tau)}+e^{\bar{\alpha}_{k}(\tau)}),\nonumber
\end{eqnarray}
where we have separated the contribution of real and imaginary parts. In the second term, each contribution is given by the real number
\begin{eqnarray}
e^{\alpha_{k}(\tau)}+e^{\bar{\alpha}_{k}(\tau)}=2e^{\alpha_{k}^{Re}(\tau)}\cos \alpha_{k}^{Im}(\tau),\nonumber
\end{eqnarray}
thus leading to a real partition function which, however, is not yet granted to be also positive. Finally, we obtain
\begin{eqnarray}
Z_{\tau}=\sum_{k\in\text{blocks}}\mu_{k}e^{\alpha_{k}^{Re}(\tau)}\cos \alpha_{k}^{Im}(\tau),\nonumber
\end{eqnarray}
which contains negative terms only for eigenvalues such that $\frac{\pi}{2} + n\pi\leq \alpha_{k}^{Im}(\tau)\leq \frac{3\pi}{2} + n\pi$, with $n$ and integer number.

A similar argument can be used to show that 
\begin{eqnarray}
\tr{e^{\hat{\mathbf{J}}(\tau)} \hat{\mathbf{J}}(\tau)}&=& \sum_{k\in\text{blocks}}\mu_{k}e^{\alpha_{k}^{Re}(\tau)}\left(\alpha_{k}^{Re} \cos \alpha_{k}^{Im}(\tau)+ \right.\nonumber\\
&-& \left.\alpha_{k}^{Im} \sin \alpha_{k}^{Im}(\tau) \right),\nonumber
\end{eqnarray}
which is still a real number. Focusing only on the eigenvalues with imaginary parts ($\alpha_{k}^{Im}(\tau)\neq 0$), we obtain a non-negative sum under the condition
\begin{eqnarray}
\frac{\alpha_{k}^{Re}(\tau)}{\alpha_{k}^{Im}(\tau)} \geq \tan \alpha_{k}^{Im}(\tau), \qquad \alpha_{k}^{Im}(\tau)\neq \frac{\pi}{2} + n\pi 
\end{eqnarray}
whereas for $\alpha_{k}^{Im}(\tau)= \frac{\pi}{2} + n\pi$ the resulting term is non-negative if $\alpha_{k}^{Re}(\tau)\geq 0$.

Note that the above conditions are stricter than what it is actually needed to ensure that the partition function and the entropy are non-negative real numbers.

\section{Weak asymmetry approximation.}

\begin{eqnarray}
    S_{ij}=\frac{1}{2}(H_{ij}+H_{ji}), \quad A_{ij}=\frac{1}{2}(H_{ij}-H_{ji}).\nonumber 
\end{eqnarray}
Let us consider the case where the asymmetric part is much sparser than the symmetric one, allowing one to approximate it as a perturbation $\delta \hat{\mathbf{S}}$ of the symmetric component. Under which condition this approximation is valid? 

Let us introduce the Frobenius norm of an operator $\hat{\mathbf{X}}$ as $||X||=\sqrt{\sum\limits_{ij} |X_{ij}|^2}$: we require that $||\hat{\mathbf{H}}||\simeq ||\hat{\mathbf{S}}||$ or, equivalently, that $||\hat{\mathbf{S}}+\hat{\mathbf{A}}||\equiv ||\hat{\mathbf{S}}||+\varepsilon$, with $\varepsilon \ll 1$. By neglecting higher-order terms in $\varepsilon$, such a condition reduces to
\begin{eqnarray}
    \varepsilon \simeq \frac{\frac{1}{2}\sum_{ij}|A_{ij}|^2 + \sum_{ij}|S_{ij}||A_{ij}|}{\sqrt{\sum_{ij}|S_{ij}|^2}} \ll 1.
\end{eqnarray}

Let us use a mean-field approximation of the quantities in the numerator and the denominator of the above expression. First, let us remind that empirical complex networks are rather sparse, i.e. the density of their edges scales as $N^{-\gamma}$, $N$ being the size of the system and with $\gamma \approx 1$~\cite{busiello2017explorability}. Equivalently, one can write that the number of edges in empirical networks follows the scaling law $E\approx c~N^{\gamma}$, with $c>1$. It follows that $\langle |X_{ij}| \rangle \simeq |\bar{x}|c N^{\gamma-2}$, where $|\bar{x}|$ is a real non-negative number. It follows that the above condition for $\varepsilon$ leads to
\begin{eqnarray}
\frac{1}{2}\frac{|\bar{a}|^2}{|\bar{s}|} + |\bar{a}| \ll \frac{1}{N^{\gamma-1}c}\nonumber 
\end{eqnarray}

Considering the experimental value $\gamma\approx 1$, and remembering that $|\bar{a}|$ and $|\bar{s}|$ are non-negative numbers, we reach the condition
\begin{eqnarray}
    |\bar{a}|\ll -|\bar{s}| + \sqrt{|\bar{s}|^2 + 2\frac{|s|}{c}} \approx \frac{1}{c},
\end{eqnarray}
which provides a very simple and elegant condition for our weakly asymmetric approximation. 

Under the above approximation, we can now determine the contribution to the total entropy of the symmetric and asymmetric components of the control operator as follows. Let us consider the two eigenvalue problems
\begin{eqnarray}
    \hat{\mathbf{S}}_{0}\vec{\mathbf{v}}_{0i}&=&\alpha_{0i}\vec{\mathbf{v}}_{0i}\nonumber\\
    (\hat{\mathbf{S}}_{0}+\delta \hat{\mathbf{S}})\vec{\mathbf{v}}_{i}&=&\alpha_{i}\vec{\mathbf{v}}_{i},\nonumber
\end{eqnarray}
where we have indicated with $\hat{\mathbf{S}}_{0}$ the unperturbed symmetric part of the control operator $\hat{\mathbf{H}}$ and its perturbation as $\delta \hat{\mathbf{S}}=\hat{\mathbf{A}}$. Using the Rayleigh eigenvalue perturbation theory we can relate the eigenvalues of the two problems as
\begin{eqnarray}
    \alpha_{i}=\alpha_{0i} + \delta \alpha_{0i}, \quad \delta \alpha_{0i}=\vec{\mathbf{v}}_{0i}^{\top}\delta\hat{\mathbf{S}}\vec{\mathbf{v}}_{0i}.
\end{eqnarray}

From Eq.~(\ref{eq:Z-Jordan-form}) we can write the Von Neumann entropy as
\begin{eqnarray}
    \label{eq:S-Jordan-form}S_{\tau}=\log\left(\sum\limits_{k}\mu_{k}e^{\alpha_{k}(\tau)}\right) - \frac{\sum\limits_{k}\mu_{k}\alpha_{k}(\tau)e^{\alpha_{k}(\tau)}}{\sum\limits_{k\in}\mu_{k}e^{\alpha_{k}(\tau)}},
\end{eqnarray}
where the index $k$ runs over the blocks of the Jordan normal form. Let us also define $\alpha_{k}(tau)=\alpha_{0k}(\tau) + \delta \alpha_{0k}(\tau)$, with $\alpha_{0k}(\tau)=-\tau \alpha_{0k}$ and $\delta\alpha_{0k}(\tau)=-\tau \delta\alpha_{0k}$, and write Eq.~(\ref{eq:S-Jordan-form}) as $S_{\tau}=S_{0\tau}+\delta S_{\tau}$, where $S_{0\tau}$ is the entropy corresponding to density matrix obtained from the control operator $\hat{\mathbf{H}}=\hat{\mathbf{S}}$, and $\delta S_{\tau}$ is a correction term corresponding to the perturbation introduced by the asymmetry  $\hat{\mathbf{A}}=\delta\hat{\mathbf{S}}$. By neglecting higher-order perturbation terms, it is possible to show that 
\begin{eqnarray}
    \label{eq:S-perturbation}\delta S_{\tau}&=&\frac{\left(\sum\limits_{k}\mu_{k}\alpha_{0k}(\tau)e^{\alpha_{0k}(\tau)}\right)\left(\sum\limits_{k'}\mu_{k'}\delta\alpha_{0k'}(\tau)e^{\alpha_{0k'}(\tau)}\right)}{\left(\sum\limits_{k}\mu_{k}e^{\alpha_{0k}(\tau)}\right)^2} \nonumber \\
    &-& \frac{\sum\limits_{k}\mu_{k}\delta\alpha_{0k}(\tau)e^{\alpha_{0k}(\tau)}\left(\alpha_{0k}(\tau)-1\right)}{\sum\limits_{k}\mu_{k}e^{\alpha_{0k}(\tau)}}, 
\end{eqnarray}
Since $S_{0\tau}\geq 0$, let us find under which condition the Von Neumann entropy is non-physical, i.e. when $S_{\tau}<0$, corresponding to the condition $\delta S_{\tau} < - S_{0\tau}$. To this aim, let us define the non-negative quantities
\begin{eqnarray}
    Z_{0\tau}&=&\sum\limits_{k}\mu_{k}e^{\alpha_{0k}(\tau)}\nonumber\\
    \bar{\alpha}_{0}(\tau) &=& \frac{\sum\limits_{k}\mu_{k}\alpha_{0k}(\tau)e^{\alpha_{0k}(\tau)}}{Z_{0\tau}},
\end{eqnarray}
which provide the partition function of the symmetric part and the mean eigenvalue. Remembering that the perturbed value of the eigenvalues are defined as $\delta \alpha_{0k}(\tau)=-\tau\vec{\mathbf{v}}_{0k}^{\top}\delta\hat{\mathbf{S}}\vec{\mathbf{v}}_{0k}$, after some algebra we obtain the final condition
\begin{eqnarray}
     Z_{0\tau}&&\left(\log Z_{0\tau} - \bar{\alpha}_{0}(\tau)\right) \nonumber \\  &<& \sum\limits_{k}\left[ \mu_{k} e^{\alpha_{0k}(\tau)}\delta\alpha_{0k}(\tau) \left(\alpha_{0k}(\tau) - 1 - \bar{\alpha}_{0}(\tau)\right)  \right]
\end{eqnarray}

\begin{table*}
\begin{longtable}{p{2.5cm} p{4cm}  p{ 4cm} p{ 5cm}}
\textbf{Type} & \textbf{Dynamical process} & \textbf{Control Operator, $\hat{\mathbf{H}}$} & \textbf{Propagator,} $\hat{\mathbf{U}}_\tau$
\endhead
\\ \hline 
\multirow{2}{*}{\textbf{Discrete}} 
& Graph Walks & $\hat{\mathbf{A}}$    & $\frac{1}{N} \hat{\mathbf{A}}^{m} (\hat{\mathbf{A}}^{m})^{\dag} $ \\[1em]
  & Random Walks (RW) & $\hat{\mathbf{K}}^{-1}\hat{\mathbf{A}}$    & $\frac{1}{N} (\hat{\mathbf{K}}^{-1}\hat{\mathbf{A}})^{m} ((\hat{\mathbf{K}}^{-1}\hat{\mathbf{A}})^{m})^{\dag} $  \\ \hline
\multirow{2}{*}{\textbf{Continuous}}
& Diffusion & $\hat{\mathbf{K}}-\hat{\mathbf{A}}$    & $ \frac{1}{N} e^{-2\tau ( \hat{\mathbf{K}}-\hat{\mathbf{A}})}$ \\
  & Continuous RW & $\hat{\mathbf{I}}-\hat{\mathbf{K}}^{-1}\hat{\mathbf{A}}$    & $\frac{1}{N} e^{-\tau (\hat{\mathbf{I}}-\hat{\mathbf{K}}^{-1}\hat{\mathbf{A}})} e^{-\tau (\hat{\mathbf{I}}-\hat{\mathbf{K}}^{-1}\hat{\mathbf{A}})^{\dag}}$ \\
    & Consensus & $(\hat{\mathbf{I}}-\hat{\mathbf{K}}^{-1}\hat{\mathbf{A}})^{\dag}$ & $\frac{1}{N} e^{-\tau (\hat{\mathbf{I}}-\hat{\mathbf{K}}^{-1}\hat{\mathbf{A}})^{\dag}} e^{-\tau (\hat{\mathbf{I}}-\hat{\mathbf{K}}^{-1}\hat{\mathbf{A}})}$
\\ \hline 
\end{longtable}
\addtocounter{table}{-1}
\caption{\label{tab:Linear_Dynamics} \textbf{Diffusive processes on complex networks.} Let the adjacency matrix be $\hat{\mathbf{A}}$ encoding the connections between every two nodes $i$ and $j$ as $\braket{j|\hat{\mathbf{A}}}{i}$, and the degree diagonal matrix be $\hat{\mathbf{K}}$, whose $ij$--th element is $\braket{j|\hat{\mathbf{K}}}{i}=k_{i}\delta_{ij}$ with $k_{i}$ being the degree of node $i$. Above is a list of some linear dynamics mentioned in the text, and their statistical propagators.}
\end{table*}

\section{Case of Hermitian control operators in the novel perspective.} Not out of necessity but for convenience, assume the probability of perturbations is uniformly distributed over all $N$ nodes, $p_i = 1/N$ and the value of perturbation is one for all nodes $\Delta_i = 1$. If the dynamical rule is linear or linearized with a Hermitian control operator $\hat{\mathbf{H}}=\hat{\mathbf{H}}^{\dag}$, the local propagator reads
\begin{equation}
    \hat{\mathbf{U}}^{(i)}_{\tau} = e^{-\tau \hat{\mathbf{H}}} | i\rangle \langle i| e^{-\tau \hat{\mathbf{H}}} 
\end{equation}

and the statistical propagator follows
\begin{eqnarray} \label{eq:statistical_propagator_hermitian}
\hat{\mathbf{U}}_{\tau} &=& \sum\limits_{i} \frac{1}{N}  e^{-\tau \hat{\mathbf{H}}} | i\rangle \langle i| e^{-\tau \hat{\mathbf{H}}}  \nonumber \\
&=&\frac{1}{N}  e^{-\tau \hat{\mathbf{H}}} (\sum\limits_{i} | i\rangle \langle i| ) e^{-\tau \hat{\mathbf{H}}} \nonumber \\
&=& \frac{1}{N} e^{-2\tau \hat{\mathbf{H}}}
\end{eqnarray}
leading to the density matrix $\hat{\boldsymbol{\rho}}_\tau = \frac{e^{-2\tau \hat{\mathbf{H}}}}{\tr{e^{-2\tau \hat{\mathbf{H}}}}}$, which is equal to the original definition of network density matrix after reparametrization $2\tau \rightarrow \tau$. Continuous diffusion on top of undirected networks is an example of application of this subsection (See Tab.~\ref{tab:Linear_Dynamics}).

\begin{table*}
\centering
\begin{longtable}{p{3cm}  p{7cm}  p{ 7cm} }
\textbf{Dynamics} & \textbf{Equation,} $\partial_\tau x_{i} =$ & \textbf{Jacobian}
\endhead
\\ \hline 
\textbf{Biochemical} & $ F - Bx_i - R\sum\limits\limits_{j=1}^{N} A_{ij}x_i x_j$ & $-B \delta_{ik} - R[\delta_{ik}\sum\limits_{j=1}^{N} A_{ij}x^{*}_{j} + (1-\delta_{ik}) A_{ij} x^{*}_{i}   ] $
\\ \hline
\textbf{Birth-death} & $ - Bx_{i}^{b} + R \sum\limits_{j=1}^{N} A_{ij} x_{j}^{a}$ & $-Bbx_{i}^{*b-1}\delta_{ik} + RaA_{ik}x_{k}^{*a-1} $ 
\\ \hline
\textbf{Regulatory} & $-Bx_{i}^{a} + R \sum\limits_{j=1}^{N} A_{ij} \frac{x_{j}^{h}}{1+x_{j}^{h}}$ &  $-Bax_{i}^{*a-1}\delta_{ik}+RA_{ik}\frac{hx_{k}^{*h-1}}{(1+x_{k}^{*h})^{2}}$ 
\\ \hline 
\textbf{Epidemics (SIS)} & $-Bx_i + R\sum\limits_{j=1}^{N} A_{ij}(1-x_i)x_j$ & $-B\delta_{ik} + R[(1-\delta_{ik})A_{ik}(1-x_i^{*})-\delta_{ik}\sum\limits_{j=1}^{N}A_{ij}x_{j}^{*}]$ 
\\ \hline 
\textbf{Synchronization} & $w_i + R \sum\limits_{j=1}^{N} A_{ij} \sin{(x_j - x_i)}$ & $-R\delta_{ik}\sum\limits_{j(\neq i)} A_{ij}\cos{(x_j^{*}-x_i^{*})} + (1-\delta_{ik})RA_{ik}\cos{(x_{k}^{*}-x_{i}^{*})}$ 
\\ \hline 
\textbf{Mutualistic} & $Bx_i(1-x_i) + R \sum\limits_{j=1}^{N} A_{ij}x_i \frac{x_{j}^{b}}{1+x_{j}^{b}}$ & $B(1-2x_{i}^{*})\delta_{ik} + R[\delta_{ik}\sum\limits_{j=1}^{N}A_{ij}\frac{x^{*b}_{j}}{1+x^{*b}_{j}} + (1-\delta_{ik})A_{ik}x_{i}\frac{bx_{k}^{*b-1}}{(1+x_{k}^{*b})^{2}}]$ 
\\ \hline 
\textbf{Neuronal} & $-Bx_i + C \tanh{x_i} + R\sum\limits_{j=1}^{N} A_{ij}\tanh{x_j}$ & $[-B+C ~\text{sech}^{2}(x_{i}^{*})]\delta_{ik} + RA_{ik}~\text{sech}^{2}(x_{k}^{*})$ 
\\ \hline
\textbf{Noisy voter} & $A - Bx_i + \frac{C}{k_i}\sum\limits_{j=1}^{N} A_{ij}x_{j}$ & $\delta_{ik}(-B+\frac{C}{k_i}A_{ik}) + (1-\delta_{ik})\frac{C}{k_i}A_{ik}$  
\\ \hline 
\end{longtable}
\addtocounter{table}{-1}
\caption{\label{tab:NONLinear_dynamics} \textbf{Linear and nonlinear dynamics on networks.} A list of dynamical equations and their corresponding control operators is listed for biochemical, birth-death, regulatory, epidemics, synchronization, mutualistic, neuronal and voter dynamics near the steady state, each having a number of constants that can be set according to the references (See Ref.~\cite{Barzel_information_flow,Barzel_signal_propagation,Barzel_universality,Carro_voter}).}
\end{table*}

Also, quantum mechanics can be described in terms of density matrices, considering the nodes of network to represent the states of the system, by inserting the physical Hamiltonian of the system $\mathcal{\hat{H}}$ as the control operator $\hat{\mathbf{H}}=-i\hbar \hat{\mathcal{H}}$ in  Eq.~(\ref{eq:statistical_propagator_hermitian}) , which gives the propagator $\frac{\hat{\mathbf{I}}}{N}$ which is expected, because the initial state here is taken to be the maximally mixed equilibrium state, and the unitaries with the same Hamiltonian can not change it. 

Starting from other states, where $p_i \neq 1/N$, the dynamics recovers the expected quantum mechanical description. Also, if the control operator is set to be $\hat{\mathbf{H}}=-\mathcal{\hat{H}}$, the statistical propagator describes the thermalization and we recover the Gibbs state, $e^{- 2\tau \mathcal{\hat{H}}}/Z_{\tau}$ for inverse temperature $\beta = 2\tau$. It is worth mentioning that the density matrix formalism presented in this paper is not necessarily for networks and can work even for the case where vectors $|i\rangle,i=1,2,3,...$ represent an infinite continuous space. In this case, the interpretation of the signal energy is the number of particles causing the perturbation and the Von Neumann entropy quantifies the mixedness of states and diversity of system's response to such perturbations.

\section{Case of non-Hermitian control operators in the novel perspective.} The new perspective allows for deriving density matrices for non-Hermitian control parameters that have valid probabilistic interpretation (real positive spectrum) and Von Neumann entropy. Again, for convenient, assume the probability of perturbations is uniformly distributed over all $N$ nodes, $p_i = 1/N$ and the value of perturbation is one for all nodes $\Delta_i = 1$. If the dynamical rule is linear or linearized with a non-Hermitian control operator $\hat{\mathbf{H}} \neq \hat{\mathbf{H}}^{\dag}$, the local propagator reads
\begin{equation}
    \hat{\mathbf{U}}^{(i)}_{\tau} = e^{-\tau \hat{\mathbf{H}}} | i\rangle \langle i| e^{-\tau \hat{\mathbf{H}}^{\dag}} 
\end{equation}

and the statistical propagator follows
\begin{eqnarray} \label{eq:statistical_propagator_nonH}
\hat{\mathbf{U}}_{\tau} &=& \sum\limits_{i} \frac{1}{N}  e^{-\tau \hat{\mathbf{H}}} | i\rangle \langle i| e^{-\tau \hat{\mathbf{H}}^{\dag}}  \nonumber \\
&=&\frac{1}{N}  e^{-\tau \hat{\mathbf{H}}} (\sum\limits_{i} | i\rangle \langle i| ) e^{-\tau \hat{\mathbf{H}}^{\dag}} \nonumber \\
&=& \frac{1}{N} e^{-\tau \hat{\mathbf{H}}} e^{-\tau \hat{\mathbf{H}}^{\dag}}
\end{eqnarray}

Continuous approximations of Random Walks (RW), maximum entropy RW and classes of consensus dynamics on top of undirected and directed networks, and also continuous diffusion on top of directed networks are examples of application of this subsection (See Tab.~\ref{tab:Linear_Dynamics}). 

\section{Discrete dynamics.} The new perspective allows for deriving density matrices for discrete types of dynamics as well as the continuous ones. If the transition matrix governing the discrete dynamics is given by $\hat{\mathbf{H}}$, the perturbation propagation vector from node $i$ after $m$ discrete steps follows
\begin{equation}
    |\Delta \phi^{(i)}_{m} \rangle = \hat{\mathbf{H}}^{m} |i\rangle
\end{equation}
assuming the field is in void state $|\phi\rangle = 0$
initially and $\Delta_i = 1$. Assuming uniformity $p_i = 1/N$, we can calculate the statistical propagator as
\begin{equation}\label{eq:statistical_propagator_discrete}
    \hat{\mathbf{U}}_m = \frac{1}{N} \hat{\mathbf{H}}^{m}  (\hat{\mathbf{H}}^{m})^{\dag} 
\end{equation}
leading to a well-defined density matrix $\hat{\boldsymbol{\rho}}_m = \frac{\hat{\mathbf{U}}_m}{\tr{\hat{\mathbf{U}}_m}}$, that has not been explored in the original formulation. A generalization for temporal networks is straightforward, having a different transition matrix $\hat{\mathbf{H}}^{(\gamma)}$ for each snapshot of the network indicated by index $\gamma$. Discrete graph walks and random walks on directed and undirected networks are examples of application of this subsection (See Tab.~\ref{tab:Linear_Dynamics}). 

\section{Nonlinear dynamics.} The new perspective allows for derivation of valid density matrices for nonlinear dynamics, as well as linear ones. For simplicity, we indicate the amount of field on top of node $i$ at temporal scale $\tau$ as $x_i = x_i(\tau) = \langle i | \phi_\tau \rangle$. 

Let us consider a generic dynamical system, defined on the top of a network $G$ of size $N$, defined by the equation
\begin{eqnarray}\label{eq:nonlinear}
\dot{\mathbf{x}}(t)=\mathbf{F}[\mathbf{x}(t),t] + \mathbf{\Sigma}[\mathbf{x}(t),t],
\end{eqnarray}
where $\mathbf{x}(t)\in \mathbb{R}^{N}$ denotes the system state, $\mathbf{F}$ denotes (non)linear deterministic functions, possibly different for each component of the state, which also accounts for the structural coupling between systems' units, and $\mathbf{\Sigma}$ denotes stochastic functions.

One can use Eq.~(\ref{eq:local_propagator}) to calculate the statistical propagator from Eq.~(\ref{eq:statistical_propagator}), for any arbitrary initial condition $x_i(0)$ and any arbitrary perturbation probabilities $p_i$. Depending on the problem, it is often prefered to find the statistical propagator where intital state is the void state $\mathbf{x}(0)=0$ or the steady state $\partial_\tau \mathbf{x}(0) = 0$. Here, we provide a list of deterministic nonlinear equations governing biochemical, birth-death, regulatory, epidemics, synchronization, mutualistic, neuronal and voter dynamics (See Tab.~\ref{tab:NONLinear_dynamics}). 

To simplify the derivation of the corresponding density matrices, in the following, we show how to linearize them near the steady state of the underlying dynamical process. For simplicity, let us focus our derivation on autonomous systems with no stochastic component. Let us assume that, under some constraints, the system of equations admits a solution $\mathbf{x}^{\star}$ corresponding to a stable or a metastable state, i.e. $\mathbf{F}[\mathbf{x}^{\star}]=0$ for a sufficient amount of time or at $t\longrightarrow \infty$, and where we are limiting to autonomous systems. Around such a state, we can study the leading-order expansion around the perturbation defined by $\delta\mathbf{x}(t)=\mathbf{x}(t)-\mathbf{x}^{\star}$, leading to the dynamical system
\begin{eqnarray}
\label{eq:deltax}\delta\dot{\mathbf{x}}(t)\simeq\mathbf{\hat{J}}_{F}\delta\mathbf{x}(t), 
\end{eqnarray}
where $\mathbf{\hat{J}}_{F}$ 
denotes the Jacobian matrix of functions $\mathbf{F}$
in $\mathbf{x}^{\star}$. In the following, we can consider that overall Jacobian matrix as the control operator $-\mathbf{\hat{H}}=\mathbf{\hat{J}}_{F}$. This way, we can analytically derive the control operators and density matrices for biochemical, birth-death, regulatory, epidemics, synchronization, mutualistic, neuronal and voter dynamics near the steady state (See Tab.~\ref{tab:NONLinear_dynamics}).


\section{Subadditivity of the Von Neumann entropy in the novel perspective.} Since density matrices in the new formalism are defined in a way to always satisfy the mathematical criteria--- i.e., being positive semi-definite, Hermitian with trace one--- they are naturally subadditive, according to the original definition in quantum mechanics. In other words, if two networks have density matrices $\hat{\rho}_{A}$ and $\hat{\rho}_{B}$, and their totall density matrix is indicated by $\hat{\rho}_{AB}$, that reduces to $\hat{\rho}_{AB}=\hat{\rho}_{A} \otimes \hat{\rho}_{B} $ if the two systems have no correlations, it is guaranteed that the summation of their entropies is not smaller than their totall entropy $\mathcal{S}_{A}+\mathcal{S}_{B}\geq \mathcal{S}_{AB}$.

Recently, a new definition of subadditivity has been introduced for networks~\cite{de2016spectral}. This criterion is useful to compare the entropy of two networks of the same size, with adjacency matrices $\hat{A}_{A}$ and $\hat{A}_{B}$, with their aggregate network having the adjacency matrix $\hat{A}_{AB}=\hat{A}_{A}+\hat{A}_{B}$. Here, we first review the derivation for the original formulation and show under what conditions it is satisfied. Then, we generalize to the new formalism.

Note that for the control operator $\hat{\mathbf{H}}$, the density matrix in the previous framework reads $\hat{\boldsymbol{\rho}}=e^{-\tau \hat{\mathbf{H}}}/Z_\tau$. Assume the control operators of two networks of the same size respectively read $\hat{\mathbf{H}}_{A}$ and $\hat{\mathbf{H}}_{B}$. Also assume that their combined control operator is given by $\hat{\mathbf{H}}_{AB}=\hat{\mathbf{H}}_{A}+\hat{\mathbf{H}}_{B}$. The relative entropy between the combined system and the first network is given by $D(\hat{\boldsymbol{\rho}}_{AB}|\hat{\boldsymbol{\rho}}_{A})=-\mathcal{S}_{AB} + \tau \tr{\hat{\mathbf{H}}_{A}\hat{\boldsymbol{\rho}}_{AB}}+ \log{Z_{A}}$ and the relative entropy between the combined system and the second network is given by $D(\hat{\boldsymbol{\rho}}_{AB}|\hat{\boldsymbol{\rho}}_{A})=-\mathcal{S}_{AB} + \tau \tr{\hat{\mathbf{H}}_{B}\hat{\boldsymbol{\rho}}_{AB}}+ \log{Z_{B}}$. Note that relative entropy is non-negative and for any pair of valid density matrix $\hat{X},\hat{Y}$ it is given that $D(\hat{X}|\hat{Y})\geq 0$. Also, when $\hat{\mathbf{H}}$ and $\hat{\boldsymbol{\rho}}$ are positive semidefinite, it can be shown that $\tr{\hat{\mathbf{H}}\hat{\boldsymbol{\rho}}}\geq 0$, following the Cholesky factorization. Finally, we assume that $\log{Z_{AB}}\geq 0$, a condition that is bound to be satisfied in cases like diffusion dynamics where $\hat{\mathbf{H}}$ has at least one 0 in the spectrum. Therefore, a summation of all such non-negative terms must be non-negative:
\begin{eqnarray}\label{eq:subadd}
D(\hat{\boldsymbol{\rho}}_{AB}|\hat{\boldsymbol{\rho}}_{A}) &+& D(\hat{\boldsymbol{\rho}}_{AB}|\hat{\boldsymbol{\rho}}_{B})
+ \tau  \tr{\hat{\mathbf{H}_{A}}\hat{\boldsymbol{\rho}}_{A}} \nonumber \\ 
&+& \tau \tr{\hat{\mathbf{H}}_{B}\hat{\boldsymbol{\rho}}_{B}} + \log{Z}_{AB} \geq 0.
\end{eqnarray}

From here, given that $\mathcal{S} = \tau \tr{\hat{\mathbf{H}}\hat{\boldsymbol{\rho}}} + \log{Z} $, it can be shown that $\mathcal{S}_{A}+\mathcal{S}_{B}\geq \mathcal{S}_{AB}$. 

It is important to note that aggregate subadditivity is proved only if all the above criteria are valid and the control operator for the aggregate network reads $\hat{\mathbf{H}}_{AB}=\hat{\mathbf{H}}_{A}+\hat{\mathbf{H}}_{B}$. For instance, in case of the combinatorial Laplacian $\hat{\mathbf{H}}=\hat{\mathbf{D}} - \hat{\mathbf{A}}$, where the degree of each node in the aggregate network is equal to the summation of its degree in each of the two networks, the diagonal matrix corresponding to the aggregate reads $\hat{\mathbf{D}}_{AB}=\hat{\mathbf{D}}_{A}+\hat{\mathbf{D}}_{B}$ and, of course, the adjacency matrix corresponding to the aggregate matrix follows $\hat{\mathbf{A}}_{AB}=\hat{\mathbf{A}}_{A}+\hat{\mathbf{A}}_{B}$. Therefore, the condition for the proof presented above is satisfied: $\hat{\mathbf{H}}_{AB}=\hat{\mathbf{H}}_{A}+\hat{\mathbf{H}}_{B}$. However, for other types of dynamics such as random walks $\hat{\mathbf{H}}=\hat{\mathbf{I}} - \hat{\mathbf{A}}\hat{\mathbf{D}}^{-1}$, the condition is not satisfied and, therefore, the aggregate subadditivity is not guaranteed.

Similarly, in case of the novel perspective presented in the paper, we can use the above formula to check if a dynamical process satisfies the aggregate subadditivity. For this reason, we use continuous dynamics with statistical propagator given by $\hat{\mathbf{U}}_\tau = e^{-\tau \hat{\mathbf{H}}} e^{-\tau \hat{\mathbf{H}}^{\dag}}  $. Here we use Baker-Campbell-Hausdorff formula

\begin{eqnarray}
 e^{-\tau\hat{\mathbf{H'}}} =&& e^{-\tau\hat{\mathbf{H}}}e^{-\tau\hat{\mathbf{H}}^{\dag}}\nonumber \\
\hat{\mathbf{H'}} =&& \hat{\mathbf{H}} + \hat{\mathbf{H}}^{\dag} - \frac{\tau}{2} [\hat{\mathbf{H}},\hat{\mathbf{H}}^{\dag}] + \frac{\tau^{2}}{12} [\hat{\mathbf{H}},[\hat{\mathbf{H}},\hat{\mathbf{H}}^{\dag}]] \nonumber \\ &-&\frac{\tau^{2}}{12} [\hat{\mathbf{H}}^{\dag},[\hat{\mathbf{H}},\hat{\mathbf{H}}^{\dag}]]+...
\end{eqnarray}

to simplify the appearance of the statistical propagator and better integrate with the above derivations. Here, similarly to the original framework, if the logarithm of the combined partition function is non-negative $\log{Z}_{AB} = \tr{e^{-\tau\hat{\mathbf{H'}}_{AB}}}\geq 0$, the operators $\hat{\mathbf{H}}'_{A}$ and $\hat{\mathbf{H}}'_{B}$ are positive semidefinite, and the operator $\hat{\mathbf{H}}'$ corresponding to the aggregate reads $\hat{\mathbf{H}}'_{AB}=\hat{\mathbf{H}}'_{A}+\hat{\mathbf{H}}'_{B}$, the aggregate subadditivity is satisfied. An straightforward example is diffusion governed by the combinatorial Laplacian $\hat{\mathbf{H}}=\hat{\mathbf{D}}-\hat{\mathbf{A}}$. In this case, since the control operator is hermitian $[\hat{\mathbf{H}},\hat{\mathbf{H}}^{\dag}]=0$, the operator  $\hat{\mathbf{H}}'$ can be written simply as $\hat{\mathbf{H}}'=2\hat{\mathbf{H}}$ and the aggregate subadditivity is satisfied, through Eq.~\ref{eq:subadd}.

\bibliography{biblio}

\end{document}